\documentclass{ws-rv10x7}
\usepackage{ws-rv-van}     
\usepackage{ws-rv-thm}     
\usepackage{subfigure}

\makeindex

\begin{document}

\newcommand{\vb}{{\bf v}}
\newcommand{\qb}{{\bf q}}
\newcommand{\pP}{{\cal P}}
\newcommand{\sigb}{\boldsymbol{\sigma}}

\addtocounter{chapter}{13}

\chapter[From viscous fluids to elastic solids]{From
  viscous fluids to elastic solids:\\ A perspective on the glass
  transition\label{ch1}}

\author[A. Zippelius and M. Fuchs]{A. Zippelius$^1$ and M. Fuchs$^2$}

\address{$^1$ University G\"ottingen, D-37077 G\"ottingen, Germany\\
annette@theorie.physik.uni-goettingen.de\\
$^2$ University of Konstanz, D-78457 Konstanz, Germany\\
matthias.fuchs@uni-konstanz.de}

\begin{abstract}
  A theory for the non-local stress  in 
  liquids  captures the crossover
  from viscous to elastic correlations upon supercooling.
  It explains the emergence of long-ranged stress fields
  in glass which originate from the
  coupling of shear stress to transverse deformations. 
  The Goldstone mode in colloidal glass is shown to be diffusive. 
\end{abstract}

\body

\section{Introduction}\label{sec1}

The most prominent distinction between a fluid and a glass is the
response to a static shear stress: A fluid continues to flow as long
as the shear is applied and hence its response is characterized by a
finite shear viscosity.  On the other hand, a glass or an amorphous
solid displays a finite deformation in response to a small applied
shear and is thus characterized by a finite elastic resistivity to shear
deformations. One of the first to capture this fundamental difference
was Maxwell~\cite{Maxwell}, who suggested that the transition from viscous to elastic
behaviour is due to the divergence of a (single) relaxation time
$\tau$. He proposed the following simple relation between a shear
stress, e.g. $\sigma_{xy}$, and the corresponding velocity gradient
\begin{equation}\label{eq1}
(\partial_t+1/\tau)\sigma_{xy}=\mu\; 
\partial_y v_x
\end{equation}
in terms of a the shear modulus, $\mu$, encoding the elasticity.
In the solid $\tau$ is infinite, implying an elastic response
$\sigma_{xy}=\mu \;\partial_y u_x$ in terms of the displacement, $u_x$, which enters Eq.~\eqref{eq1} via its time derivative, $v_x=\partial_t u_x$. In the fluid, $\tau$
is finite, so that in the hydrodynamic limit $\sigma_{xy}=\eta \partial_y v_x$ with the shear viscosity given by $\eta=\mu\tau$.

What is missing? We know from linear elasticity theory that stresses are long ranged in solids: A localised shear strain, $\epsilon_{xy}$, generates far away stresses according to~\cite{Landau,Eshelby1957,Picard2004,Lemaitre2015}
\begin{eqnarray}\label{Eq:Eshelby}
    \sigma_{xy}({\bf {r}})&=&2\mu \int d^3r'G({\bf {r}}-{\bf {r}'})
                              \epsilon_{xy}({\bf {r}'})\\ \notag
    G({\bf {r}})&=&\frac{3}{4\pi r^7}\big(r^2(x^2+y^2)-10x^2y^2\big)
                    \propto r^{-3}
   \end{eqnarray}
   Note that the Green function, $ G({\bf {r}})$, is not only
   long-ranged, but also anisotropic, even though we have specialised
   to isotropic solids, such as glasses or amorphous solids.
The basic questions we want to address in this paper are the following:
How do long range stress correlations build up at the glass transition?
  Are there precursors in the supercooled liquid?
  Can we formulate a unified hydrodynamic theory of liquids and glasses?

To answer these questions, we have computed the correlations of the {\bf local} shear stress fluctuations~\cite{Maier2017,Maier2018}
\begin{equation}\label{eq2}
  C({\bf q})=\frac{n}{k_BT}\langle \sigma_{xy}({\bf-q},t)\sigma_{xy}({\bf q}) \rangle\;
\end{equation}  
where the homogeneity of the system is conveniently exploited by  Fourier transforming.  The microscopic stress tensor that enters Eq.~\eqref{eq2} is taken from Irving and Kirkwood~\cite{IrvingKirkwood} which reads for the potential contribution:
\begin{equation}
\sigma_{\alpha\beta}({\bf r})=\sum_{<j,k>}
  \frac{{\bf r}_{jk}^{\alpha}{\bf r}_{jk}^{\beta}}{r_{jk}}U'(r_{jk})a
  \int_0^1ds\delta({\bf r}- {\bf r}_{k}-s{\bf r}_{jk}).
\end{equation}
The shortcoming of a single relaxation time approximation \`{a} la Maxwell
is the neglect of slow dynamics in the local stress fluctuations, which cannot be captured by a single relaxation time and which will ultimately give rise to nontrivial stress correlations.
  
\section{Emergence of long range stress correlations}\label{sec2}

The obvious candidates for slow relaxation in a Newtonian fluid are
the conserved densities of particle number, momentum and energy. For
simplicity, we consider an incompressible, isothermal system and refer
to the literature~\cite{Maier2018,Vogel2019} for the general case. The only conserved field is
thus the transverse momentum or velocity, defined as
${\bf v}^\perp({\bf q})={\bf q}\times({\bf q}\times{\bf v}({\bf
  q}))/q^2$ with
${\bf v}({\bf q})=\frac{1}{\sqrt{N}}\sum_{i=1}^N e^{i{\bf q}{\bf
    r}_i(t)}{\bf v}_i$.

In the hydrodynamic limit, its correlation function
$\langle \vb^\perp(\qb,t)^*\; \vb^\perp(\qb) \rangle = ({\bf 1} - \frac{\qb\qb}{q^2})\; {\rm K}_q(t)$ describes diffusive momentum transport at long wavelengths; in the fluid phase the relaxation rate $\propto \eta q^2$ diverges as the wavenumber goes to zero, reflecting the conservation law. This behaviour is captured in the following representation~\cite{Kadanoff} of the Laplace transform $K_q(s) = \int_0^\infty\!\!dt\, e^{-st} K_q(t)$
\begin{equation}
{\rm K}_q(s) =  \frac{k_BT/m}{s + \frac{q^2}{mn} {\rm G}_0(s)} 
\end{equation}
in terms of a generalised shear modulus ${\rm G}_0(s)$. The main advantage
of this representation is that it guarantees the correct treatment of
the conservation law and allows for simple approximations of the
modulus. For example, a single relaxation time approximation \`{a} la Maxwell,
${\rm G}_0(s)=(\mu\tau)/(1+s\tau)$, reproduces the diffusion of transverse momentum in the fluid ($\tau$ finite)
and (undamped) transverse sound modes in the solid ($\tau$ infinite).

The conservation of transverse momentum gives rise to a slow component
also in the relaxation of the stress
correlation~\cite{Maier2017,Maier2018}, which can be isolated with
help of the Mori-Zwanzig formalism~\cite{Latz}. A projection operator
$\pP=\frac{m}{k_BT}\, \vb^\perp(\qb) \rangle\!\cdot\! \langle
\vb^\perp(\qb)^\ast$ captures the overlap between the transverse
momentum and a fluctuation of the shear stress. Application of $\pP$
to the stress correlation, $ C({\bf q})$, yields the desired
decomposition~\cite{Maier2017,Maier2018}
\begin{equation}
C(\qb,s) ={\rm G}_0(s)- \big((q_x^2+q_y^2)-4\frac{q_x^2q_y^2}{q^2} \big) \frac{({\rm G}_0(s))^2}{n k_BT}  \; {\rm K}_q(s)
\end{equation}
into a hydrodynamic contribution and local dynamics entailed in ${\rm G}_0(s)$. 
In a straightforward generalization of the Maxwell model, we use a single relaxation time approximation for 
${\rm G}_0(s)=(\mu\tau)/(1+s\tau)$, allowing for a divergence of the structural relaxation time $\tau$ at the glass transition.

What are the predictions of the generalised Maxwell Model? First, the
result of Maxwell is reproduced for the global stress in a fluid,
$C({\bf {q=0}},s)=\eta$ with shear viscosity $\eta=\mu\tau$. Second,
stress correlations are strongly {\bf anisotropic} in the isotropic
fluid, characterized by finite $\tau$:
\begin{equation}\label{Eq:stresscorr}
C({\bf {q}},s)=\eta-\big(q_x^2+q_y^2-\frac{4q_x^2q_y^2}{q^2} \big)
\frac{\eta^2}{nms+\eta q^2}.
\end{equation}
Local stresses do not decay quickly, but display long-lived diffusive
behaviour. The distance to the glass transition is controlled by
$\tau$, or equivalently $\eta$, which is known to increase
dramatically as the glass transition is approached. Here, we follow
Maxwell and consider an ideal glass transition with a true divergence
of $\tau$. The increasingly slow dynamics implies increasingly {\bf
  long-ranged} stress correlations as the glass transition is
approached, i.e. $\tau\to \infty$. The spatial extent of stress correlations is quantified by a correlation length $\xi$ which diverges as the glass transition is approached, $\xi^2=\mu\tau^2/(mn)$.
For high frequencies $s\tau\gg 1$, the fluid supports tranverse sound,
as one would expect.
 Third, the glass (characterized by infinite $\tau$) exhibits
a time-persistent part of the stress correlation:
\begin{equation}
\lim_{s\to 0}s C({\bf {q}},s)=C_{\infty}({\bf {q}})=
\frac{4q_x^2q_y^2+q_z^2q^2}{q^4}\mu.
\end{equation}
The glass resists static shear
deformations and the above static correlations are equivalent to
Eshelby`s response function in Eq.~(\ref{Eq:Eshelby}). Shear deformations in the glassy
phase are propagating sound modes as reflected in the connected
correlation:
$ C({\bf {q}},s)-C_{\infty}({\bf {q}})= \frac{\mu s}{s^2+q^2c^2}$ for
$q_x=0$ and with speed of sound $c^2=\mu/(mn)$. Fourth, with the above definition of the correlation length
we can write the stress correlation in scaling form:
$C({\bf {q}},t)=\mu{\cal {F}}(t/\tau,{\bf {q}}\xi)$, accounting for the stress fluctuations in the fluid, at the glass transition, and in the glassy phase.
This scaling function is shown in Fig.~1 in the $q_x,q_y$ plane for
several values of rescaled time $t/\tau$. The left panel corresponds
to a fluid with $t/\tau=10$, the middle one depicts the transition
region with $t/\tau=2$ and the right panel corresponds to the glassy
regime with $t/\tau=0.1$. The anisotropy is clearly visible in the four
fold symmetric pattern for all three cases. The corrections to the
simple Maxwell model vanish along the diagonal
($q_x=q_y=q/\sqrt{2}$). Here the anisotropic terms in Eq.~\ref{Eq:stresscorr} vanish and
$ C({\bf {q}},t)$ is constant. Along the axis (either $q_x=0$ or
$q_y=0$), where force-correlations are tested~\cite{Evans1981}, the deviations from Maxwell are strongest with (undamped) transverse sound modes in the solid.

\begin{figure}
\begin{minipage}[b]{0.29\textwidth}
 \includegraphics[width=1.0\textwidth] {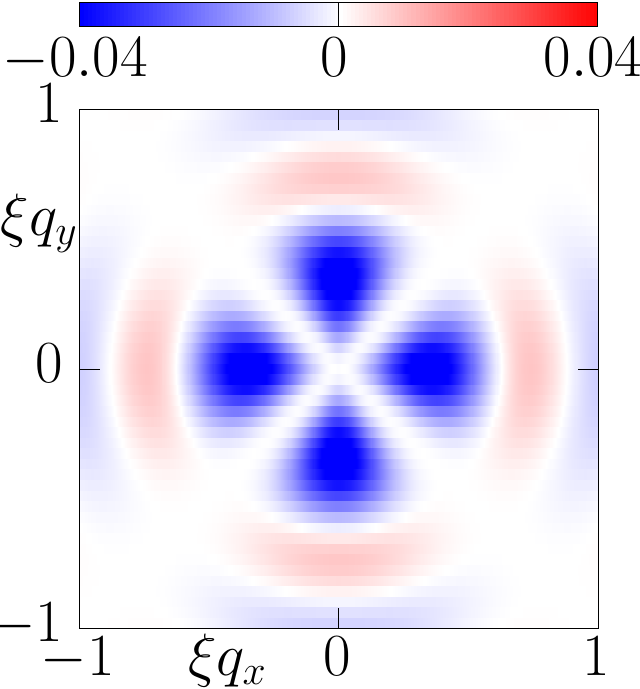}
\end{minipage}$\quad$
\vspace{-0.5cm}
\begin{minipage}[b]{0.3\textwidth}
 \includegraphics[width=1.0\textwidth]{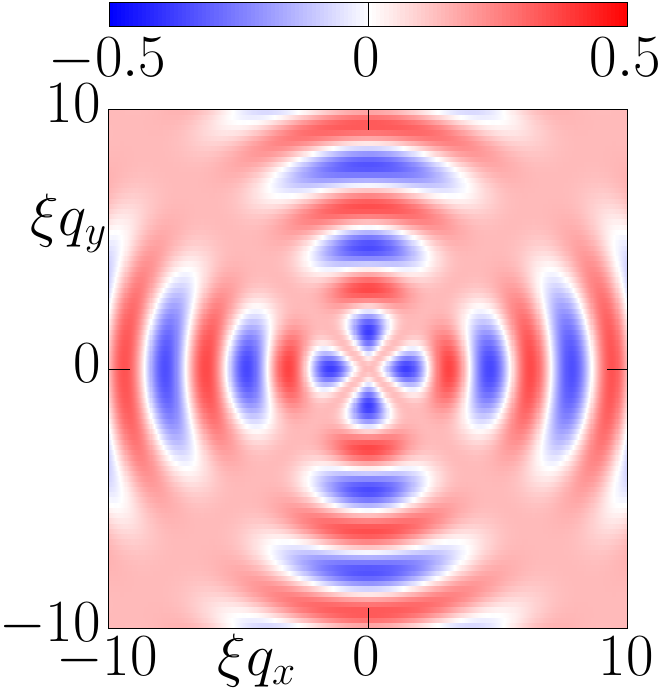}
\end{minipage}$\quad$
\begin{minipage}[b]{0.32\textwidth}
\includegraphics[width=1.0\textwidth]{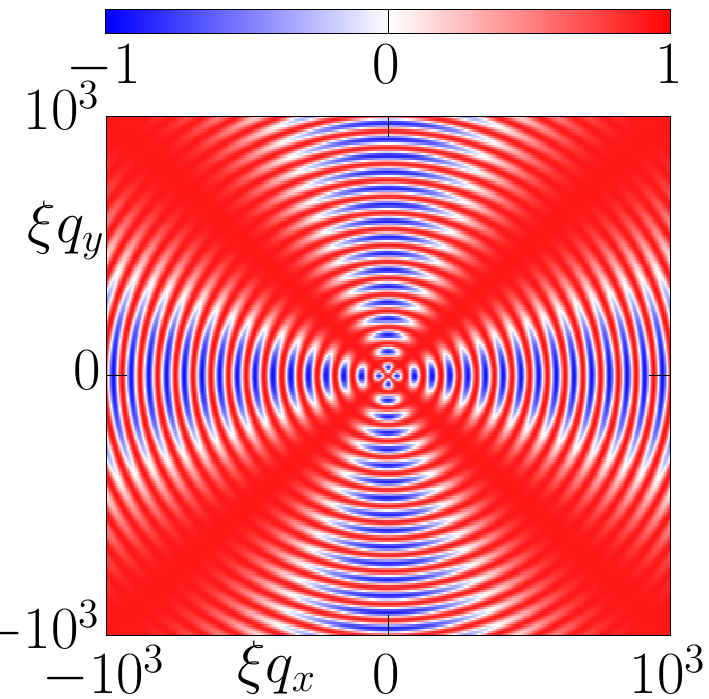}
\end{minipage}\\
\caption{Shear stress correlations $C({\bf {q}},t)/\mu={\cal {F}}(t/\tau,{\bf {q}}\xi)$ in the generalized Maxwell model for fluid ($t/\tau=10$),  viscoelastic  ($t/\tau=2$), and solid  ($t/\tau=0.1$) state points, from left to right; from Ref.~\cite{Maier2017}. }
\end{figure}

\section{Goldstone modes of a colloidal glass}

The momentum of colloidal particles in suspension is not conserved in
contrast to the Newtonian fluid discussed above~\cite{Dhont}.
The interaction with the solvent is conveniently approximated by a constant friction coefficient $\zeta_0$, ignoring hydrodynamic interactions. Consequently, velocity correlations 
\begin{equation}\label{K_coll}
{\rm K}_q(s) =  \frac{k_BT/m}{s + \frac{q^2}{mn} {\rm G}_0(s)+n\zeta_0}
\end{equation}
decay on microscopic timescales in the hydrodynamic regime and, due to symmetry,
there are no conserved fields which could give rise to slow shear stress fluctuations. However, we expect that correlations of
the transverse displacement, ${\bf u}^\perp$, are long ranged also in colloidal 
glasses. To capture these long range fluctuations in a
unified hydrodynamic theory of supercooled liquids and glasses including suspensions,
we separate the dynamics of stress fluctuations into a part in the
subspace of ${\bf u}^\perp$, or rather its time derivative
${\bf v}^\perp=\partial_t{\bf u}^\perp$, and the rest. In other words
we use the same decomposition~\cite{Vogel2019} as for the Newtonian
case with however different velocity correlations (Eq.~\ref{K_coll}).

The most surprising results of the generalised Maxwell model refer to the colloidal glass. The static elasticity, as described by $C_{\infty}({\bf q})$, is the same as for the Newtonian model as one would expect. However, frequency dependent transverse deformations propagate diffusively
\begin{equation}\label{Goldstone}
  C({\bf {q}},s) - C_{\infty}({\bf {q}})= \mu\; \big(\frac{q_x^2+q_y^2}{q^2}-
  4\frac{q_x^2q_y^2}{q^4}  \big)\; \frac{\zeta_0}{s\zeta_0+q^2 \mu/n}
\end{equation}
We identify this diffusive mode with the Goldstone excitations of an
amorphous solid. Localisation of the particles implies a spontaneous
breaking of the translational symmetry of the system. In contrast to
crystalline systems, the symmetry is restored on a macroscopic level,
because the particles are localised at random positions~\cite{Mukhopadahyay2004}. Nevertheless,
a uniform translation of all particles leaves the energy invariant
and the energy of an almost uniform translation goes to zero as the
wavelength of the perturbation grows.  In a phenomenological approach,
we start from the elastic free energy $F=\mu/2\int
d^dq\,q^2\,{\bf u}^{\perp}(\bf{q})\cdot{\bf u}^{\perp}(-\bf{q})$ 
and assume purely relaxational dynamics
\begin{equation}
  n  \zeta_0 \partial_t{\bf u}^{\perp}(\bf{q})=-\frac{\delta F}{\delta {\bf u}_{\perp}(-\bf{q}) }=-\mu \; q^2 \;{\bf u}^{\perp}({\bf q}).\nonumber
\end{equation}
The relaxation of ${\bf u}^{\perp}(\bf{q})$ is diffusive in perfect agreement with the diffusive pole $s=-q^2\mu/(n\zeta_0)$ observed in Eq.~(\ref{Goldstone}).\\

Precursors of this diffusive mode can be observed in supercooled
colloidal suspensions, which are well described by Eq.~(\ref{Goldstone}) in the large damping limit, ignoring inertial terms. The frequency dependent spectra are strongly anisotropic in ${\bf q}-$ space, as already observed for the Newtonian case. Choosing ${\bf q}$ along one of the axis, say ${\bf q}=(q,0,0)$ we find:

\noindent
\begin{minipage}[b]{0.54\textwidth}
  \begin{eqnarray}
 C({\bf q},s=-i\omega)&=C^\prime({\bf q},\omega)-iC^{\prime\prime}({\bf q},\omega)\nonumber\\   
C^{\prime
  \prime}(q,\omega)&=\mu\frac{\omega\tau(1+q^2\xi^2)}{(q^2\xi^2+1)^2+\omega^2\tau^2}\nonumber\\
C^{\prime}(q,\omega&=\mu\frac{\omega^2\tau^2}{(q^2\xi^2+1)^2+\omega^2\tau^2}\nonumber  
  \end{eqnarray}
  Here we have introduced the correlation length $\xi^2=\mu\tau/(n\zeta_0)$, which diverges as $\xi\propto \sqrt{\tau}$ in contrast to the Newtonian case.
  For sufficiently large $q\xi>1$, the peak in the loss spectrum is located at
  $\omega\tau\sim q^2\xi^2$.
\end{minipage}
\hfill
\begin{minipage}[b]{0.44\textwidth}
\includegraphics[width=1.0\textwidth]{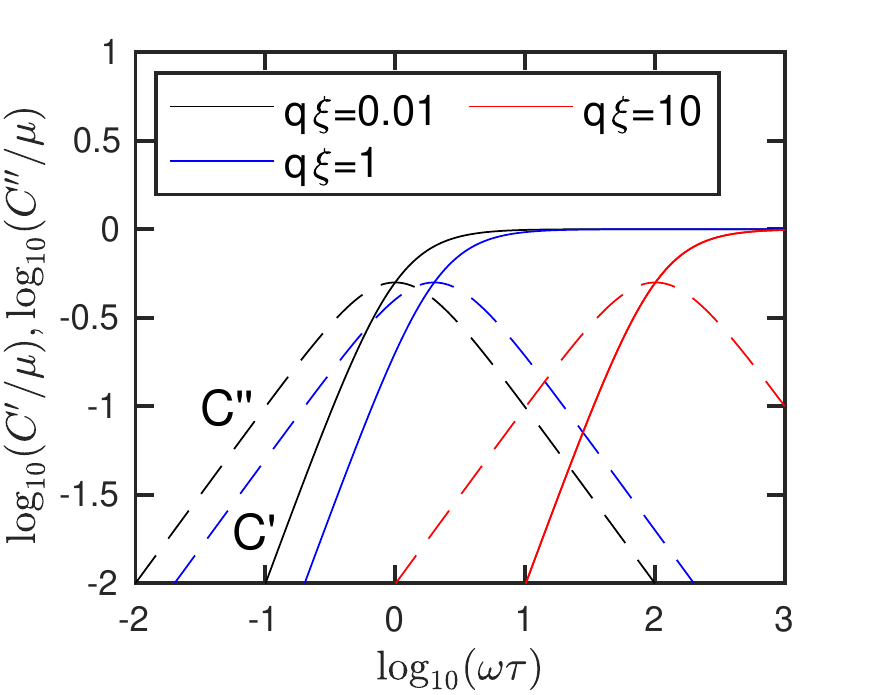}
{\small Fig. 2. Shear stress spectra; from Ref.~\cite{Vogel2019}. }
\end{minipage}\\

\section{Conclusions}

We have computed the nonlocal stress correlations for both, Newtonian
as well as colloidal fluids. A generalised Maxwell model connects the emergence of
long range stress correlations in the viscous fluid to the elasticity
of glasses. In Newtonian fluids, the range of stress correlations is
charcterized by a correlation length $\xi\propto \tau$, which grows
approaching the glass transition, whereas in colloidal fluids
$\xi\propto\sqrt{\tau}$.  In the colloidal glass, transverse
deformations propagate diffusively in contrast to transverse sound in
the Newtonian case.

The nonlocal stress correlations can equivalently be computed from
purely hydrodynamic
considerations~\cite{Maier2017,Maier2018,Vogel2019,Vogel2020},
decomposing the flow field into the externally imposed flow and a
fluctuating part. The response function to the externally imposed
flow is related via the fluctuation-dissipation theorem to the
correlation function, discussed here.  Several
extensions of the simple model are possible; we have already analysed
compressible systems~\cite{Maier2018,Vogel2019,Klochko2018}, including
longitudinal sound. Another extension refers to a computation of the
generalised modulus, going beyond the single relaxation time
approximation.

\section*{Acknowledgments}
All the results presented here were done in collaboration with Manuel
Maier and Florian Vogel, whose contributions are gratefully
acknowledged.

\bibliographystyle{ws-rv-van}
\bibliography{ws-rv-sample}

\end{document}